\pdfoutput=1
\documentclass[10pt,twocolumn]{confpaper}

\graphicspath{ {./figures/} }

\date{}
\title{\ttlfnt{
\TITLE}}
\author{
\aufnt{\AUTHORS} \\
\affaddr{Princeton University} \\
{\normalsize \url{http://interconnection.citp.princeton.edu/}}
\vspace*{-0.1in}
}

\begin{document}

\maketitle


\begin{sloppypar}

\thispagestyle{version}
\begin{abstract}
The rapidly evolving nature of interconnection has sparked an increased interest
in developing methods for gathering and collecting data about 
utilization at interconnection points. One mechanism, developed by DeepField Networks, allows
Internet service providers (ISPs) to gather and aggregate utilization
information using network flow statistics, standardized in the Internet
Engineering Task Force as IPFIX. This report (1)~provides an overview of
the method that DeepField Networks is using to measure the utilization of
various interconnection links between content providers and ISPs or
links over which traffic between content and ISPs flow; and (2)~surveys
the findings from five months of Internet utilization data provided by
seven participating ISPs---Bright House
Networks, Comcast, Cox, Mediacom, Midco, Suddenlink, and Time Warner
Cable---whose access networks represent about 50\% of
all U.S. broadband subscribers.

We first discuss the problem of interconnection and utilization at
interconnection points. We then discuss the basic operation of the
measurement capabilities, including the collection and aggregation of
traffic flow statistics (i.e., IPFIX records), providing an assessment
of the scenarios where these aggregate measurements can yield accurate
conclusions, as well as caveats associated with their collection.  We
assess the capabilities of flow statistics for measuring utilization,
and we discuss
the capabilities and limitations of the approach
the aggregation techniques that the ISPs use both in providing data to
us, and that we apply before making the data public. 

The dataset includes about 97\% of the paid peering, settlement-free
peering, and ISP-paid transit links of each of the participating ISPs.
Initial analysis of the data---which comprises more than 1,000 link
groups, representing the diverse and substitutable available
routes---suggests that many interconnects have 
significant spare capacity, that this spare capacity exists both across
ISPs in each region and in aggregate for any individual ISP, and that the aggregate utilization across
interconnects is roughly 50\% during peak periods.

\end{abstract}

\ifthenelse{\equal{\onlyAbstract}{no}}{

\section{Introduction}\label{sec:intro}

As traffic demands increase due to the rise of
large asymmetric traffic flows such as
 video
streaming,
interconnection arrangements must evolve to meet these new demands. The nature, causes,
and location of Internet congestion has spawned contentious debate over
the past two years. End users have become increasingly invested in this
topic as well, although they have sometimes conflated the issues of Internet
congestion with other concerns about the prioritization of Internet
traffic.

Discussion about interconnection can and should be better informed by
accurate, up-to-date information about where capacity bottlenecks
exist. Unfortunately, until now, 
data about traffic utilization at Internet interconnection points has
been hard to come by, due to confidentiality and business
constraints. This opacity has led users, policymakers, and researchers
to resort to techniques that attempt to isolate congestion using
end-to-end probes~\cite{coates2001network,www-ndt,www-mlab}, which
nonetheless still leave significant uncertainty 
about where congestion may be occurring.  

One of the biggest barriers to furthering this debate is the lack of
clear data on this problem. As the Internet pioneer David Clark recently
said, ``An issue that has come up recently is whether interconnection
links are congested. The parties who connect certainly know what's going
on, but that data is generally not disclosed. The state of those links
matters to a lot of people ... and there have been some
misunderstandings around congestion and interconnection
links''~\cite{clark:nanog66}.   

To help shed light on this important issue, ISPs have provided
unprecedented data around the state of interconnection links. This data
yields some information concerning the utilization of network ports that
face each network's ``peers'' (i.e., the networks that each ISP connects
to directly). This data aims to illuminate the utilization properties of
each network’s externally facing switch ports and ascertain whether each
collection of ports between a given ISP and its respective neighboring
network is uncongested. Although this data cannot, by itself, tell the
complete story about the location of congestion along end-to-end
Internet paths, it can tell us a lot about where congestion is not
occurring.  

Each participating ISP has provided information about its interconnection to
neighboring networks (e.g., ISPs, content providers) in each region, as well as the capacity of each interconnect.
The data that participating ISPs provide account for about 97\% of links
from all participating ISPs in any given month; the only links that are
missing from the dataset are those where the measurement infrastructure
has not yet been deployed.  This information offers sufficient
information to ascertain the capacity of each interconnect between an
ISP and neighboring networks. Given this information, we can compare
this provisioned capacity against traffic statistics for traffic that
traverses each of these network ports and compare the measured
utilization to the provisioned capacity to achieve an estimate of
utilization. The ability to perform this analysis depends on the ability
to collect accurate, utilization measurements. Section~\ref{sec:method}
discusses the collection method.

Ideally, the information we would be able to see the utilization
and capacity for each individual port, for every ISP---in such a
scenario, comparing utilization to provisioned capacity would be
straightforward. Of course, the practical realities are more
complicated: even the {\em existence} of an individual interconnection is
typically considered proprietary, not to mention the business agreement
surrounding that interconnection, as well as the capacity and
utilization of the interconnection. As a result, the data that the ISPs
provide aggregates sampled flow statistics across link groups 
in each region, providing a high-level picture of capacity and
utilization per region and ISP, as well as how this utilization
fluctuates over time. The traffic
flow statistics, based on IPFIX~\cite{rfc7011} and
collected by DeepField Networks~\cite{www-deepfield} represents
utilization information that is collected at the interconnection points,
thus providing a more direct indication of the utilization information
at interconnection points.

The data does have some limitations that make it inappropriate for
answering certain questions about utilization. First, it is sampled,
which makes it difficult to answer certain types of questions about flow
size distributions, characteristics of small flows, and utilization by
application.  Second, to preserve proprietary information, the data is
aggregated and anonymized, preventing conclusions about utilization at
specific interconnection points. Yet, the data illuminates
interconnection capacity and utilization at many levels. Throughout this
report, we are careful to highlight conclusions that we can and cannot
make with the data that the participating ISPs have
provided. Based on feedback from other experts, we have also iterated on
the data that the ISPs have agreed to release, resulting in a careful
balancing act between preserving proprietart information and revealing
information about utilization at interconnection 
points that can inform ongoing debates. 
Subsequent sections of this
report provide additional detail on the method used to collect and
report this data, as well as what we can and cannot conclude from the
data that the ISPs have agreed to provide. 

This paper reflects our current understanding of capacity and
utilization at interconnection points; we recognize that the dialog
surrounding interconnection is ongoing.  As a resource to interested
parties---and to promote further academic research in this field, 
we will periodically update the 
findings and data from this project on the project
website~\cite{www-citp-interconnection}. In cooperation with the
participating ISPs, we will annually assess whether the project remains
relevant as Internet interconnection evolves. We also expect
potential future opportunities to correlate this data with
performance measurements from other sources, which will shed more
light into the relationship between utilization at 
interconnection and end-to-end performance.

The rest of this paper is organized as follows.
Section~\ref{sec:related} describes related work and analysis
techniques. Section~\ref{sec:method} describes the measurement techniques and
data, as well as the effects of
various phenomena such as sampling on the accuracy of the collected
data. Section~\ref{sec:applicability} discusses where the measurements
  from this study can (and cannot) be applied. 
  Section~\ref{sec:findings} describes the findings from a preliminary
  analysis of the data collected as part of the
  project. Section~\ref{sec:conclusion} concludes with a summary and
  suggestions for possible next steps.
\section{Related Data and Analysis Techniques}\label{sec:related}

In this section, we briefly outline existing attempts to measure both
end-to-end performance of Internet paths and infer
congestion along these paths (and at interconnects) using these
datasets.  
All of these techniques and approaches involve inference based on
measurements from end hosts, as opposed to direct measurements of
utilization at the interconnect. As a result, public data about
utilization and capacity at the interconnects---which this project
provides for the first time---fills a significant
gap concerning our visibility into the current state of utilization at
interconnects.  

\subsection{Measurements from End-Hosts}

A common approach to performing Internet performance measurements is to
actively send test traffic along end-to-end Internet paths and observe
the performance characteristics of those paths. For example, one might
perform test uploads or downloads from an end-user device (laptop,
phone, home gateway device) and measure the time to transfer a certain
number of bytes. Similarly, it is possible to measure end-to-end latency
or packet loss along these end-to-end paths, as well as to measure how
these characteristics may vary in response to additional load on the
network. 

\paragraph{Measurement Lab.} The Measurement Lab~\cite{www-mlab}
operates global server infrastructure for conducting throughput
measurements from various endpoints, using pre-approved measurement
tools. Measurement Lab (MLab)  limits the tools that can perform
throughput measurements against their servers due to the fact that
server bandwidth is a limited resource. One of the tools that has
permission to measure against this infrastructure is
BISmark~\cite{sundaresan2011,www-bismark} , which we 
describe in more detail below. Other tools for measuring mobile
performance (e.g., MobiPerf) exist. The tool that perhaps offers the
most comprehensive data from the project is the Network Diagnostic Tool
(NDT), which we also describe in more detail below.

\paragraph{Network Diagnostic Tool (NDT).} The network diagnostic tool (NDT)~\cite{www-ndt}
performs throughput tests; users run NDT from end-hosts,
which measure throughput to a corresponding server. One version of the
tool runs as a Java applet from a web browser. Measurement Lab runs a
version of the Java applet from its website that measures throughput to
the collection of deployed Measurement Lab servers around the world,
using geolocation to map the client to a nearby NDT server for the
purposes of the throughput test (the accuracy of a TCP throughput test
depends on measuring throughput to a nearby server, since TCP throughput
is inversely proportional to round-trip latency). 
NDT also forms the basis of well-known measurement efforts, such as the
Internet Health Test. 
Unfortunately, MLab's NDT
test setup is known to be inaccurate due to its use of only a single thread to
measure TCP throughput, which our previous work shows can significantly
underestimate the throughput of the link~\cite{sundaresan2011}. Additionally, NDT
provides no mechanism for locating a throughput bottleneck along an
end-to-end path.

\paragraph{BISmark.} The Broadband Internet Service Benchmark
(BISmark)~\cite{sundaresan2011,www-bismark} 
project runs custom throughput, latency, and packet loss measurements
from home routers that run
OpenWrt. The project has been
collecting 
performance data from home networks since 2011; at its peak, the project
was collecting data from about 400 home networks in more than 30
countries. Currently, about 70 home routers are actively reporting
measurements. The project was the first research effort to explore the
means of measuring access-network throughput and latency of a broadband
access network’s access link using direct measurements. All of the data
is publicly available, both through a web portal, and via direct
download in XML format. The BISmark measurement techniques perform
end-to-end measurements against deployed servers and do not attempt to
draw inferences or conclusions about congestion at interconnect. The
BISmark project produced the first published research paper that
documented interconnection congestion at many interconnects that
occurred in March 2014; because the measurements were end-to-end, they
manifested as pronounced increases in latency along specific end-to-end
paths between home Internet subscribers and the M-Lab
servers. Subsequent work, which we describe in the next section, has
followed up on this effort in more detail. 

\paragraph{FCC Measuring Broadband America Reports.} The FCC’s Measuring
Broadband America project~\cite{www-mba} produces periodic reports using similar measurements
as the BISmark project, albeit with a much larger deployment. Their
reports are less frequent (typically once per year), as opposed to
BISmark's ``real time'' visualizations of throughput, latency, and packet
loss. The techniques are similar (some of them, such as the
throughput test and the Web performance test, were co-designed), as are
the servers against which the home network gateways perform measurements
(i.e., both perform throughput measurements against the Measurement Lab
servers). Similarly, the project does not provide any mechanism for
directly measuring congestion at interconnection points; the only
performance measurements that the devices can perform are end-to-end
performance measurements. 

\subsection{Measuring Interconnect Performance}

Because the above tools can only measure from end-host vantage points,
they do not provide direct information about utilization or congestion
at interconnection points. Because congestion manifests as an
increase in latency, the measurement techniques that we have discussed
above can often detect congestion along an Internet path. Yet, detecting
congestion at a particular interconnection point is difficult to do with
these types of measurements. We discuss various other methods to
indirectly infer or directly measure congestion at an interconnection
point below.

\subsubsection{Indirect: Tomography \& Round-Trip Latency}

A general measurement approach sometimes referred to as {\em network
tomography} attempts to use a combination of performance measurements
along different end-to-end Internet paths to infer specific links where
congestion or failures may be occurring~\cite{coates2001network}. The intuition is quite simple:
Given ``simultaneous'' measurements of two end-to-end Internet paths that
may share one or more links, if one end-to-end path experiences symptoms
of congestion (i.e., an increase in latency) whereas a second end-to-end
path does not, then we can infer that the congestion must be occurring
on the portion of the second path that is not common with the first
path. One can generalize this to N end-to-end paths; the hope is there
is some set of end-to-end paths in the measurement infrastructure such
that each link could be isolated.  

Unfortunately, it is difficult to obtain a comprehensive set of vantage
points in practice because most end-to-end paths will share more than
one interconnection point or link in common. For example, in an M-Lab
report released in 2014~\cite{mlab-congestion}, many of the end-to-end paths between NDT
vantage points and the M-Lab servers could (and likely do) share
multiple end-to-end links along the path---not only the interconnection
point (where the report implies congestion is taking place) but also
other links along the path (e.g., links within transit providers).  The
second scenario is a
distinct possibility that previous reports have outlined in
detail~\cite{ftt-congestion-2015}, and it would be na\"{i}ve to suggest that
these 
measurements conclusively pinpoint congestion at interconnection
points. Worse yet, providers can (and have) gamed these active
measurement techniques by prioritizing probe traffic~\cite{kilmer2014}.

Another approach, proposed by CAIDA~\cite{www-caida}, is to use traceroutes to
discover an end-to-end 
path and subsequently send latency probes to either side of an
interconnect. While this approach is more direct than network
tomography, the approach entails significant shortcomings, which are
outlined in detail in CAIDA's own report. Among the limitations are the
difficulty in accurately identifying points of interconnection points
along and end-to-end traceroute, as well as the fact that increases in
latency might be occurring along reverse paths, as opposed to the
forward path that the probes are attempting to measure. 

\subsubsection{Direct: Packet Capture and SNMP}

An alternative method for directly gathering information about the traffic that
passes through a network is via packet traces.  Packet traces capture
what is effectively a recording of every packet that traverses a
particular interface. When packet trace collection is configured, an
administrator may capture the complete packets, the first bytes of each
packet, or simply the ``headers'' or metadata for each packet. Packet
capture provides complete timing information about the arrival of
individual packets and the header information on individual packets
(including the TCP window size), and can as such be used to compute or
infer properties of traffic flows including jitter, packet loss, and
instantaneous throughput.  

It would be beneficial to have better information about latency and
packet loss to assess the congestion status of a particular flow, as
well as what might be causing poor user quality of experience, but these
types of conclusions typically require gathering packet-level
statistics.  Methods such as deep-packet inspection are typically not
practical at large, high-throughput interconnection points; these
methods tend to be costly to deploy, and they produce more data than can
be reasonably backhauled to a data-center for post hoc analysis.
Additionally, typically gathering fine-grained, packet-level information
is not tenable at high packet rates, so gathering traffic flow
statistics must often suffice.  Traffic flow statistics are quite a bit
more coarse, because they only provide information about the number of
bytes and packets transferred over the duration of the
flow record.  Accordingly, although these types of methods may be
appropriate for certain types of analysis pertaining to quality of
experience, security, or other network management tasks, the question of
utilization of interconnects is best answered today with flow-level
statistics (e.g., IPFIX) or SNMP counters.  Traffic flow statistics can
represent traffic statistics on a per-flow basis as opposed to SNMP byte
counts, which only represent total interface utilization counts. SNMP
statistics are thus more coarse for many purposes, because they do not represent the
utilization of specific flows and are polled at relatively infrequent
intervals.

\paragraph{AT\&T/DirectTV Merger Analysis.} The Cooperative Association
for Internet Data Analysis~(CAIDA)~\cite{www-caida} is participating in an ongoing
consultation with the Federal Communications Commission~(FCC) concerning the
performance metrics that should be reported in conjunction with the
merger between AT\&T and DirectTV, to ensure that the network is
offering suitable performance to all traffic flows~\cite{fcc-att}.  Reporting on
performance at interconnection points is a condition of the
merger. The FCC appointed CAIDA as an independent measurement expert
(IME) to recommend a set of metrics that should be included in these
reports on interconnection performance; the recommendations suggest
including metrics such as packet loss, latency, and jitter of flows at
the interconnection point, as proxies for congestion. It is unclear at
this time what information, if anything, will be publicly reported. 

In contrast, this study only
reports on utilization, but these measurements are made public. Although
these recommendations suggest that 
utilization need not be a proxy for congestion (indeed, it may be
possible to engineer an interconnect to run at 95\% capacity or higher),
it is also worth noting that {\em none} of these metrics tell the
complete story. Similarly, packet loss may increase as a result of
active queue management or traffic shaping, as opposed to
congestion. Similarly, latency or jitter are only indirect proxies for
congestion.

\section{Measurement and Data}\label{sec:method}

We now describe the data that each ISP provides concerning the
utilization of each network port, and how this data is sampled and
aggregated.  

Sampling and aggregation can affect the accuracy of the
resulting measurements, and we discuss the effects of sampling and
aggregation later in this section.  In addition to discussing the
methods that the ISPs use, we also describe alternative approaches to
measuring network utilization and the advantages and drawbacks of each
method.

\subsection{Traffic Flow Statistics and Utilization}
A common method for gathering statistics about the utilization of a
network---and the method that this project uses---is to gather what are
often referred to as ``flow statistics''; the most common version of
flow statistics is likely the IPFIX protocol (often instantiated in
Cisco products as ``NetFlow'')~\cite{rfc7011}.  Many other vendors have conformed to a
similar standard when exporting records about traffic flows.

A IPFIX record contains metadata about the flow, including the number
of bytes transferred, the number of packets in the flow, the start and
end times for the flow, and the network interface associated with the
flow.  Accordingly, the statistics in a flow record can give useful
information about the average utilization over a period of time in terms
of either bytes or packets.  For example, if the flow record has a
duration of ten seconds and reports that 1 gigabyte of traffic was
transferred during that ten seconds, then the average utilization over
that ten-second period would be 800 megabits per second (eight gigabits
per ten seconds).  The flow statistics can also be used to compute
average packet rates, in terms of packets per second, in a similar
manner.

\begin{figure}[t]
\includegraphics[width=\linewidth]{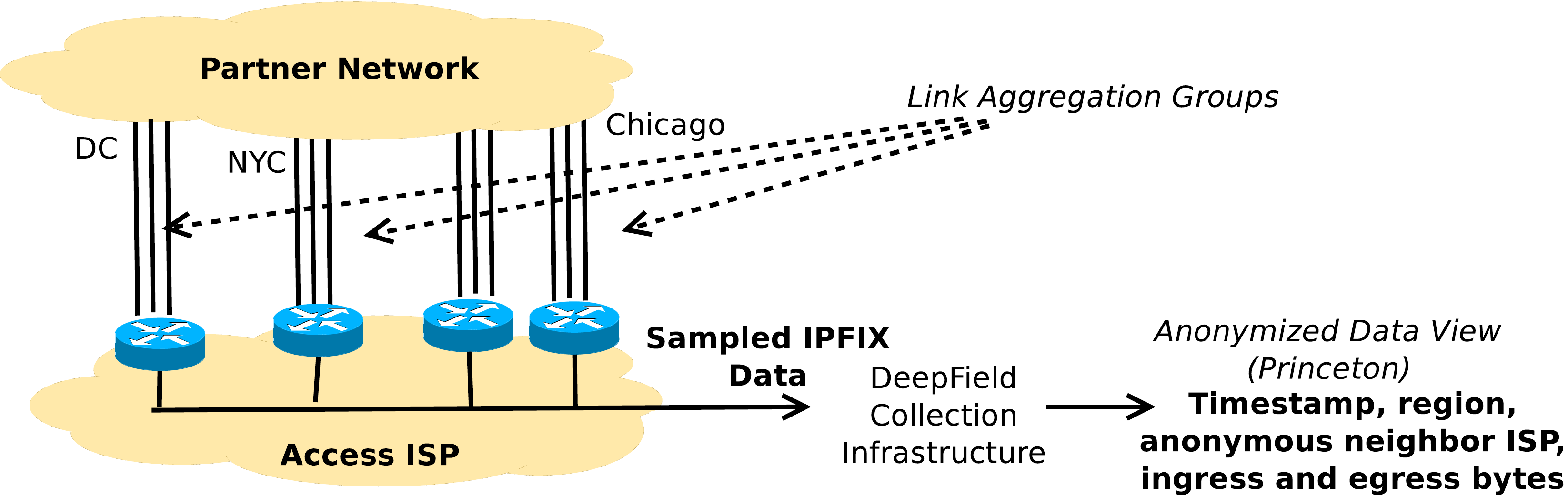}
\caption{Data collection
  infrastructure and approach.} 
\label{fig:arch}
\end{figure}

The traffic in this dataset covers interconnection points
for access ISPs that account for about 50\% of the broadband subscribers in the United
States. 
Figure~\ref{fig:arch} shows how data is collected from each
interconnection point between an access ISP and neighboring partner network.  Each participating access
ISP may connect to a partner network in multiple geographic regions. The
access ISP collects IPFIX data at each interface that interconnects with
a neighboring partner network.
The traffic statistics that each ISP reports are based on IPFIX records
that are exported at least as frequently as every 60 seconds and subsequently
aggregated across a 
link group; to protect the confidentiality of information pertaining to
usage on specific interconnects, the data is aggregated into a single link group per geographic
region. (Section~\ref{sec:aggregation} describes this approach in more
detail, and how it affects the conclusions we can draw.) The statistics
represent an aggregate that is computed based on 
the sum of peak five-minute intervals in each hour, for each {neighbor
  network, circuit group} pair.

The dataset contains about 97\% of links from all participating ISPs in
any given month; the only links that are missing from the data set are
links that are not configured in DeepField's measurement system.  All
interconnections between participating ISPs and neighboring partner
networks are private (i.e., none of the interconnections in this study
involve public IXP switch fabrics).  Each row in the dataset that CITP
receives includes the following statistics:
\begin{itemize}
\itemsep=-1pt
\item	Timestamp (representing a five-minute interval)
\item	Region (representing an aggregated link group)
\item	Anonymized partner network
\item       Access ISP
\item	Total ingress bytes
\item	Total egress bytes
\item       Capacity
\end{itemize}
\noindent
Because flows do not begin and end on discrete five-minute intervals,
each five-minute timestamp represents the sum of utilization of active
traffic flows that were active during that interval. Suppose that, at a
given time, a set of flows are active. Then, the total ingress bytes for
that five-minute interval for a single flow would be the average bitrate
for that flow over its total duration, multiplied by the amount of time
that the flow was active during the given five minute interval. The
total utilization for the link aggregation group is the sum of all such
statistics, for any flows that were active during that five-minute
interval.

\subsection{Aggregation and Load Balancing}\label{sec:aggregation}
When measuring the contribution of a traffic flow to a link's utilization,
it is also important to ensure that flows are not double counted. An
ISP's ports may be configured as a link aggregation group (we are aware
of this configuration for at least one ISP in the study).  In this ISP's case,
the router balances outbound traffic flows across the links; a single
flow always goes across a single link. The allocation of outbound
traffic flows to links is based on a hashing algorithm on the router;
given enough traffic flows, this type of load balancing typically works
well enough to balance load evenly across the available links in any
given aggregation group.  It is extremely rare for any ISP
to have multiple LAGs in a region to a given partner network.

We are cognizant of only the outbound load balance mechanisms for all of
the ISPs that contribute data; we are unaware of the traffic load
balance practices of partner networks that do not participate in the study, but, for
the purposes of assessing inbound traffic loads across links in an
aggregation group, it is likely reasonable to assume that these ISPs
also use typical load balancing practices for outbound traffic (and,
hence, we can assume a relatively uniform load balance of inbound
traffic flows for a link aggregation group). 

In networks where there exist only a small number of flows, such as in
commercial VPNs, it is possible that utilization across links might
become unbalanced, but on links carrying consumer traffic, such as those
in this study, it would be highly unusual for traffic to be unevenly
balanced across links, due to the nature of the router hashing function,
which is designed to randomly assign these flows to available links.
Thus, although traffic statistics are reported in aggregate across an
aggregation group, it is highly unlikely that we would encounter a
situation where average traffic flow statistics would report low
congestion, but some links in the aggregation group would be congested
while others would be underutilized. 

\subsection{Sampling}

IPFIX records must be sampled, meaning that the statistics in any given
record are based not on all of the packets in that flow, but rather a
random sample of the packets in that flow. In this project, ISPs report
statistics that are based on sampled IPFIX records.  Typically, IPFIX
sampling can take one of two forms: {\em random} and {\em
  deterministic}. If the sampling factor is N, then random sampling will
incorporate the statistics for any given packet with probability 1/N; on
the other hand, deterministic sampling will incorporate the statistics
based on every Nth packet deterministically.

The effects of sampling on overall traffic volume estimation bears some
discussion. Certainly, when trying to estimate certain characteristics,
such as the number of small flows that cross an interface, or the
overall distribution of flow-sizes, aggressive sampling can distort
measurement accuracy. On the other hand, estimating overall utilization
is possible in general---flows may be missed entirely, but on
average, some fraction of the small flows will be captured. Attempts to
normalize the flow sizes for small flows will result in inaccurate
estimates of the flow sizes of these small flows, but the estimates for
overall traffic volume should remain reasonably accurate. For example,
suppose that a link creates flow statistics based on a packet-sampling
rate of 1/1,000. In the extreme case, suppose that each flow is a single
packet. Then, on average, the statistics will reflect one in every
thousand flows, and attempts to normalize these statistics would result
in an estimate of one flow of 1,000 packets. Clearly, the flow-size
estimates are incorrect, but the total utilization is accurate,
on average.

The observation
that sampled IPFIX records are sufficient for aggregate capacity
utilization holds empirically, as well. We compared the SNMP byte
counters to sampled IPFIX records with a 1/8,000 sampling rate across
250 interconnect links for one of the largest participating ISPs in the
study for a single day. The mean and median of the ratio
between both metrics were both around 0.98, with a standard deviation of
0.095. As ISPs increase their sampling rates, the accuracy of IPFIX
relative to SNMP should improve further.

In conclusion, the average of sampled utilization across port groups may
underestimate utilization, and averaging across port groups may not be
able to characterize the distribution of utilization (and congestion)
across the group (e.g., some ports may be congested while others remain
uncongested). Yet, we can certainly use this data to determine with confidence
whether there exist uncongested ports in a region between a pair of
networks.

\paragraph{Sampling rates in this study.}
Most of the Internet service providers in the study report traffic flow
statistics based on a sampling rate of 1/1,000, meaning that statistics
are collected based on a sampling of every thousandth packet, on
average; all of the ISPs who are contributing data implement a sampling
rate of at least 1/8,000.  Some of the ISPs in the study use deterministic sampling
and others use random sampling; given that the goal is to estimate
capacity on links where much of the traffic flows that contribute to
congestion are large, long-running video streams---which have fairly
large packet and byte counts---neither the sampling rate nor the mode of
sampling should affect the accuracy in estimating the overall
utilization.

\subsection{Configuration and Topology}

Each participating Internet service provider (ISP) provides the
following information from configuration data, and from SNMP polling: 
\begin{itemize}
\item {\em Interconnection.} For each of an ISP's peers, the ISP's router
  configuration data provides information about which interface maps to
  each neighboring autonomous system (AS), as well as the policies
  associated with each connection, such as Border Gateway Protocol
  configuration options (e.g., local preference, and AS path
  prepending).  The router configuration also provides information such
  as the mapping of individual network interface names to the AS that
  the interface corresponds to. In this study, the next-hop AS was
  determined from BGP routing information gathered from the
  interconnection router. 

\item {\em Provisioning.} In addition to the mappings between interfaces and
  ASes that the configuration provides, SNMP polling data yields
  information about the interface capacity that is provisioned on each
  link.  
\end{itemize}

DeepField has the ability to collect this data from all routers in the
network---including those that peer directly with neighboring autonomous
systems (ASes) and those that are internal to the network. For the
purposes of this study, data from the border routers alone suffices, as
we are not concerned with internal utilization but rather only with
utilization that may occur at the edge of the network. 

\subsection{Public Use of Data}

Although the ISPs make the above data available to us, much of this data
is bound by mutual non-disclosure agreements between the ISPs and their
respective partner networks, due to the proprietary nature of
interconnections. As mentioned, both the existence and nature of any
particular interconnection is considered proprietary, as are the
decisions about where any particular ISP has a point of presence and
where any ISP tends to route different types of traffic.  These details
reflect both business strategy (e.g., provisioning), business
relationships, the source and destination of traffic demands, and
decisions about network management and operations.  We emphasize that
the restrictions on our ability to disclose data to the public result
not from a specific agreement with the ISPs but rather from the {\em
  mutual non-disclosure agreements between the ISPs and their content
  providers}, which are intended to protect both parties.

Due to the sensitive nature of much of this information, 
the public dataset reports utilization that is aggregated by region and
across at least three participating ISPs.
The
publicly released visualizations and underlying data include 
statistics about link aggregation groups, as we describe below.
The public dataset reports the following aggregate utilization statistics:
\begin{itemize}
\item For each ISP, across all interconnect links
  to all neighbor networks.
\item For each region, across all ISPs
\item Across all interconnects and all regions.
\item Across all links, both per-link and weighted by overall aggregate capacity.
\end{itemize}
\noindent
The public visualizations and underlying data, which we plan
to update monthly, reveal the following aggregate statistics and information: 
\begin{itemize}
\item Peak utilization at an interconnect, relative to total capacity, aggregated across ISPs in that region.
\item The fraction of interconnects that experience a percentage maximum
utilization, for the 95th percentile of five-minute intervals.
\item Utilization by region over time, for all regions with at least three operators.
\end{itemize}
\noindent
This level of aggregation does not make it possible to assess the
overall utilization of a particular ISP's connections to a neighboring
network, and analysis of the public data cannot show that there are no
highly utilized links. Although the private data has information about
utilization of individual interconnections, we are not permitted to
disclose statistics at this granularity; in an effort to disclose as
much information as possible, we have released certain information about
utilization of individual interconnection links, including the
distribution of utilization across these links.

 Demonstrating this result would require analysis of much more
fine-grained data.  Nonetheless, the public aggregate statistics
do provide evidence that each participating ISP and region has spare
capacity at respective interconnection points, as we discuss in more
detail in the coming sections.
\section{Limitations}\label{sec:applicability} 

In this section, we briefly discuss the applicability of the measurement
techniques for various purposes. We survey the types of conclusions
can and cannot be drawn from sampled and aggregated IPFIX measurements. 

\subsection{Limitations of Flow Statistics}

Traffic flow statistics are commonly used to estimate link utilization
for purposes of capacity estimation and planning, and for traffic
engineering purposes. Large transit provider networks commonly deploy
IPFIX across all of the routers in their networks to determine whether
certain links are overutilized.  As previously discussed, even sampled
IPFIX records can be useful for determining {\em aggregate} link
utilization.  Nonetheless, sampled IPFIX records have certain
limitations that make them inappropriate for certain types of
analysis. While these additional features would undoubtedly shed more
light on both congestion and application performance, the currently
deployed technologies do not permit these types of analyses at the
interconnection points.  The rest of the section discusses various
measurements that are not possible with the existing measurement
approach.

\paragraph{Analysis of small flows.} Due to the sampling rates of the
measurements, performing any analysis that is specific to small flows or
on the distribution of flows may not be possible. As previously
discussed, this affects our ability to analyze statistics such as
flow-size distribution but should not have any affect on our ability to
estimate utilization.

\paragraph{Timing, loss, or quality of experience.} Traffic flow
statistics also do not capture timing effects or accurate statistics
about packet loss, jitter, and so forth. Due to the lack of detailed
information that aggregate traffic flow statistics provide, inferring
properties that directly relate to user quality of experience will be
difficult with the existing dataset, given only aggregate volumes. 

\paragraph{Information about specific applications.} Additionally,
assessing the performance of any given application will be difficult
with the given dataset, since the traffic flow statistics do not have
any application-specific identifiers or other information that would
help associate the traffic with a particular end-user
application. Traffic flow statistics are gathered on
flows, which correspond to source and destination IP address and port,
as well as protocol type. Yet, this information alone does not provide
enough information to infer the application type of a flow, since
applications often share the same destination port (in particular, many
applications, including streaming video and the web, use destination
port 80). Associating performance with a particular application
will require more precise statistics, including possibly information from the
application layer or associated domain name system (DNS) lookup
information.

\paragraph{Statistics on short timescales.} The
traffic statistics represent aggregates across a group of links and
across time (typically the duration of a particular flow). As a result,
the statistics cannot capture fluctuations that may occur on short
timescales; for example, a traffic flow may send a high volume of
traffic over a relatively short interval and low volume for the
remainder of the flow duration. 
Utilization may spike on short timescales, and such spikes
would not be reflected in aggregate traffic flow statistics, since one
can really only compute an average utilization over the duration of time
that the flow record reflects. 
Because the aggregate statistics reflect
only average utilization across the duration of a flow, the statistics
will reflect these short-term fluctuations. 

\subsection{Limitations Due to Aggregation}

Even in the private dataset, statistics are reported in aggregate link
groups. In this case, any fluctuations that occur on only a single link
may not be reflected in the aggregate statistics. We previously
described assumptions about traffic load balance that suggest that
drawing conclusions based on average utilization per link is reasonable.
Additionally, short-term periods of high utilization across the entire
link group may not be evident in the data, because utilization is
reported on five-minute averages.

In the public dataset, it is possible assess the overall utilization in
some region across all ISPs and partner networks, but not for any
individual interconnection point in a region.  Similarly, it is possible
to see the aggregate utilization for any of the participating ISPs, but
not for a specific region or neighbor ISP.  As a result, the aggregates
make it difficult to drill down into the utilization between any pair of
networks, either as a whole or for any particular region.  As a result,
it is not possible to conclude that no interconnection links experience
high utilization. Because the public data shows utilization across each
ISP, we can conclude that each ISP has spare capacity---although we
cannot conclude that it has spare capacity in each region or on any
individual port.  

To mitigate concerns that result from this level of aggregation, the
public dataset also includes 95th percentile peak utilizations for all
links in the dataset, which demonstrates that most of the links in the
dataset as a whole experience low utilization, and that much of the
aggregate capacity remains under-utilized even at peak. We also show the
aggregate utilization for all ISPs in each region, which allows us to
demonstrate that each region has spare capacity; because this statistic
is aggregated across ISP, we cannot conclude that a particular ISP has
spare capacity in a region---especially to a specific neighbor.  Yet,
the our ability to show spare capacity in aggregate increase confidence
that this capacity exists, since most ISPs have significant spare
capacity at peak utilization, and most links in the dataset have spare
capacity at peak, as well.


\section{Utilization at Interconnection Points}\label{sec:findings}

In this section, we present preliminary analysis of the utilization
measurements from the interconnect groups from the participating
ISPs. We survey the capacity and utilization of each interconnect group
both overall and by region. From October 2015 through February 2016, 
aggregate interconnect capacity has been roughly 50\% utilized at peak,
and capacity has grown consistently by about 3\% monthly, or about 19\% over the
five-month period.
In the rest of
this section, we explore the utilization characteristics of these links.

\begin{figure}[t]
\includegraphics[width=\linewidth]{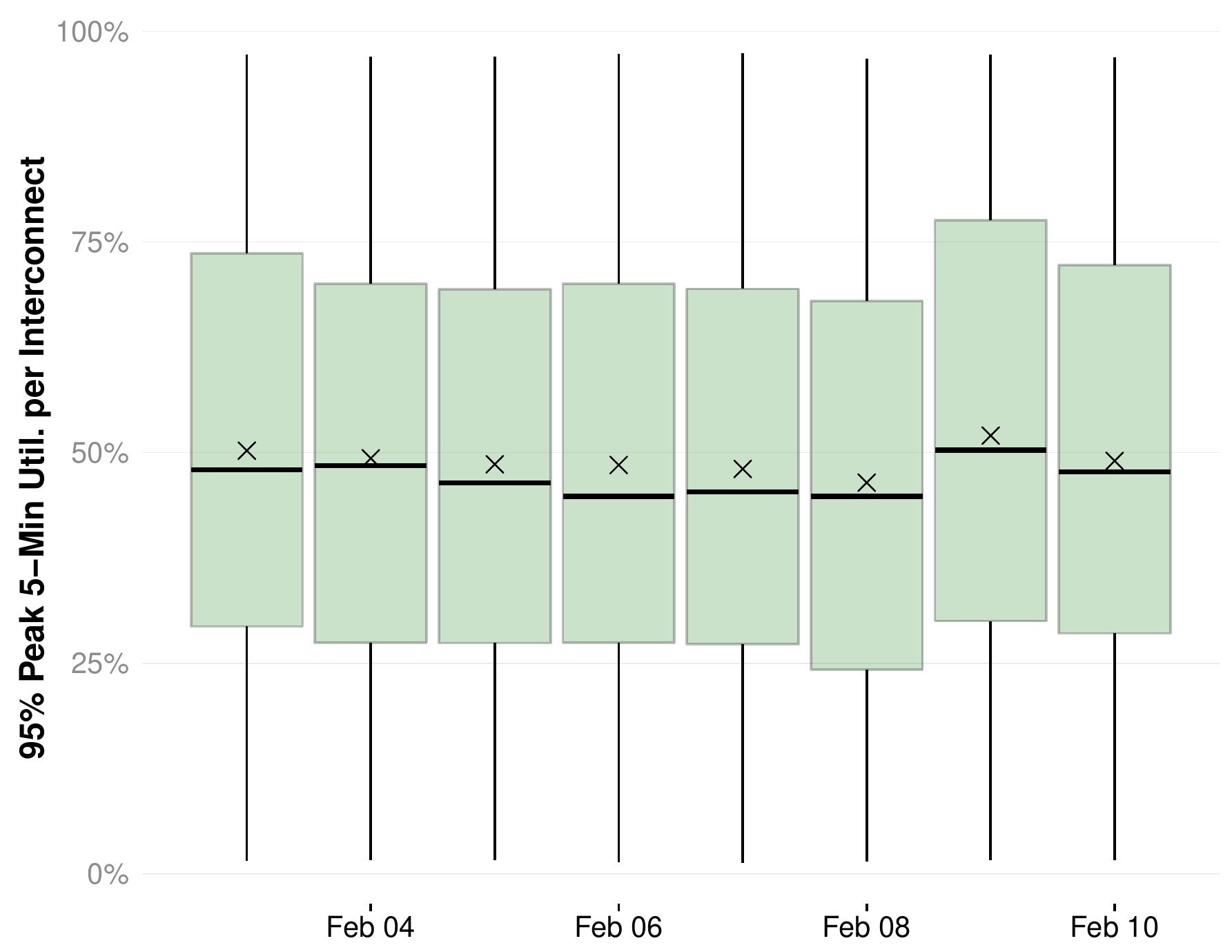}
\caption{Utilization of each interconnect group over one week in
  February 2016, normalized by
  capacity of the interconnects.} 
\label{fig:utilization-timeseries}
\end{figure}

\begin{figure}[t]
\includegraphics[width=\linewidth]{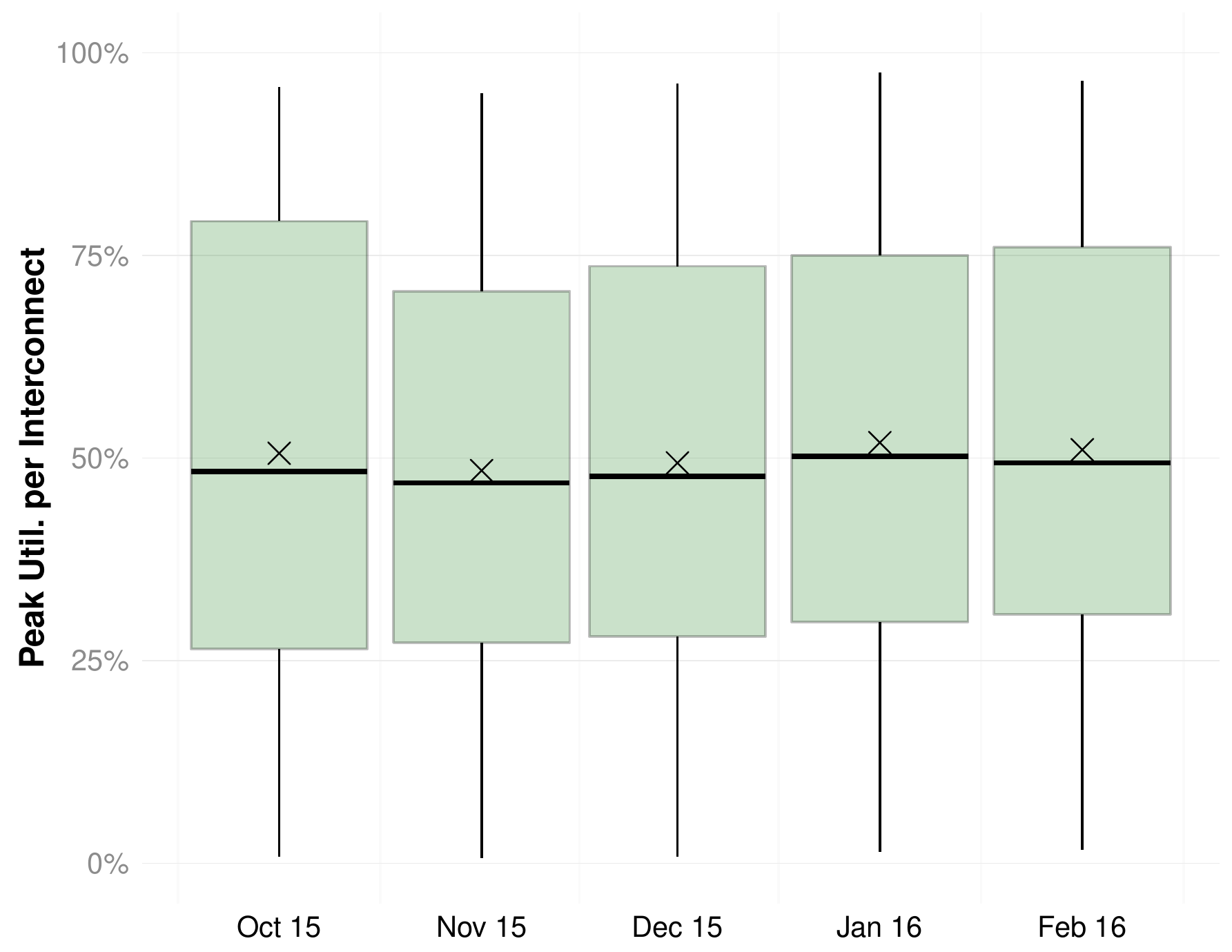}
\caption{Per-month utilization of all participating interconnects.} 
\label{fig:utilization-per-month-all}
\end{figure}

\begin{figure}[t]
\includegraphics[width=\linewidth]{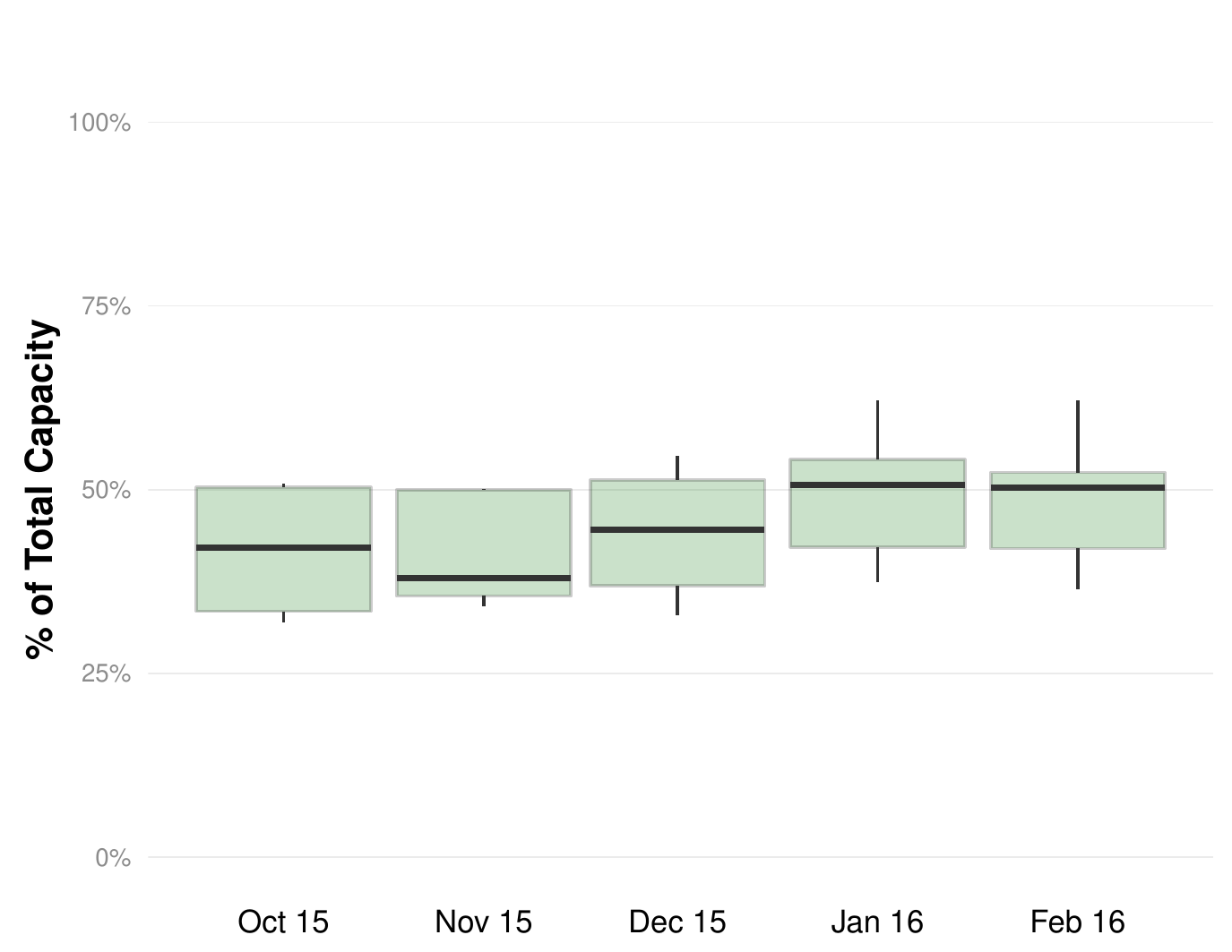}
\caption{Distribution of 95th percentile peak ingress utilization across
  all ISPs, with all ISPs equally weighted.} 
\label{fig:isp-dist}
\end{figure}

\begin{figure}[t!]
\begin{minipage}{1\linewidth}
\begin{subfigure}[b]{\linewidth}
\includegraphics[width=\linewidth]{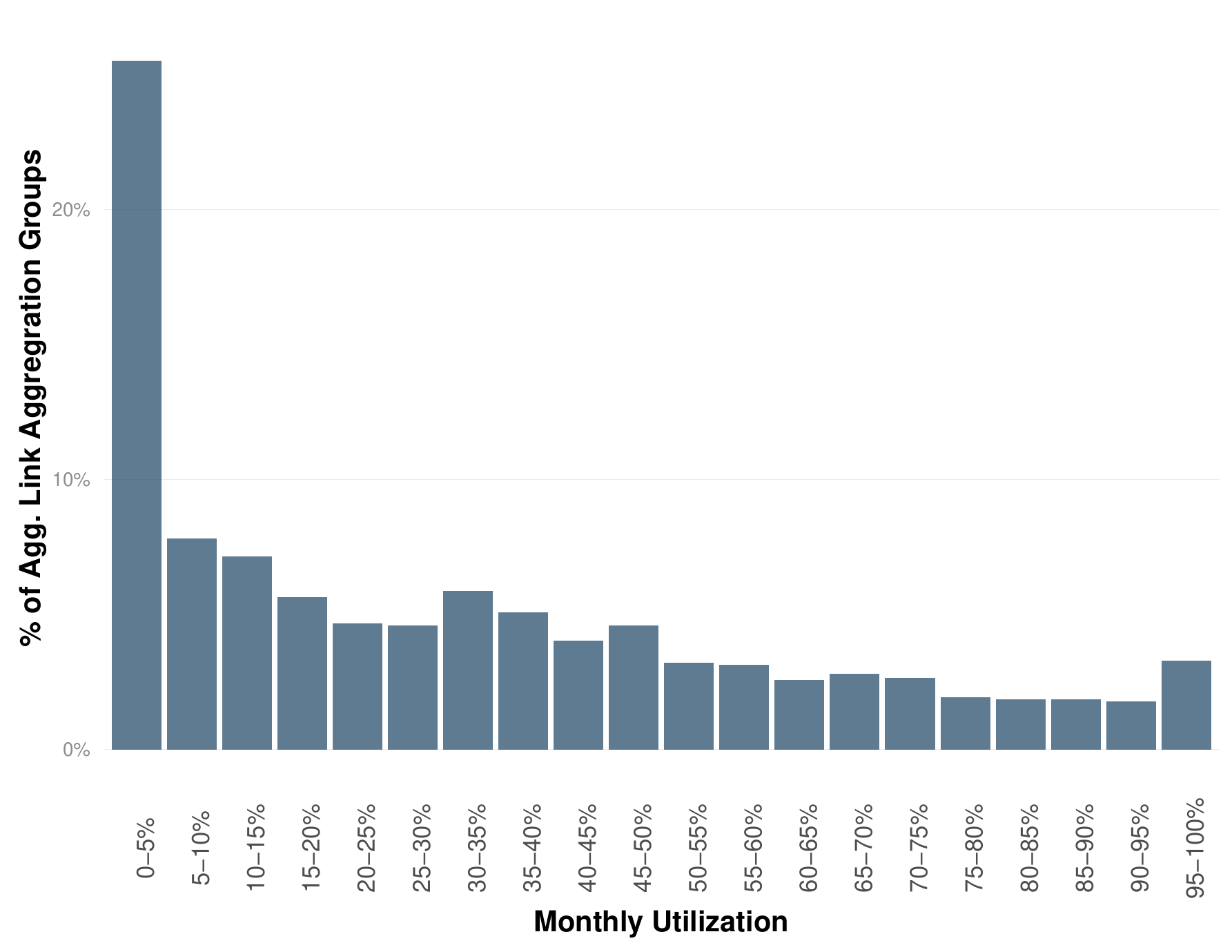}
\caption{Weighted by links.\label{fig:dist-link-peak}}
\end{subfigure} \hfill
\begin{subfigure}[b]{\linewidth}
\includegraphics[width=\linewidth]{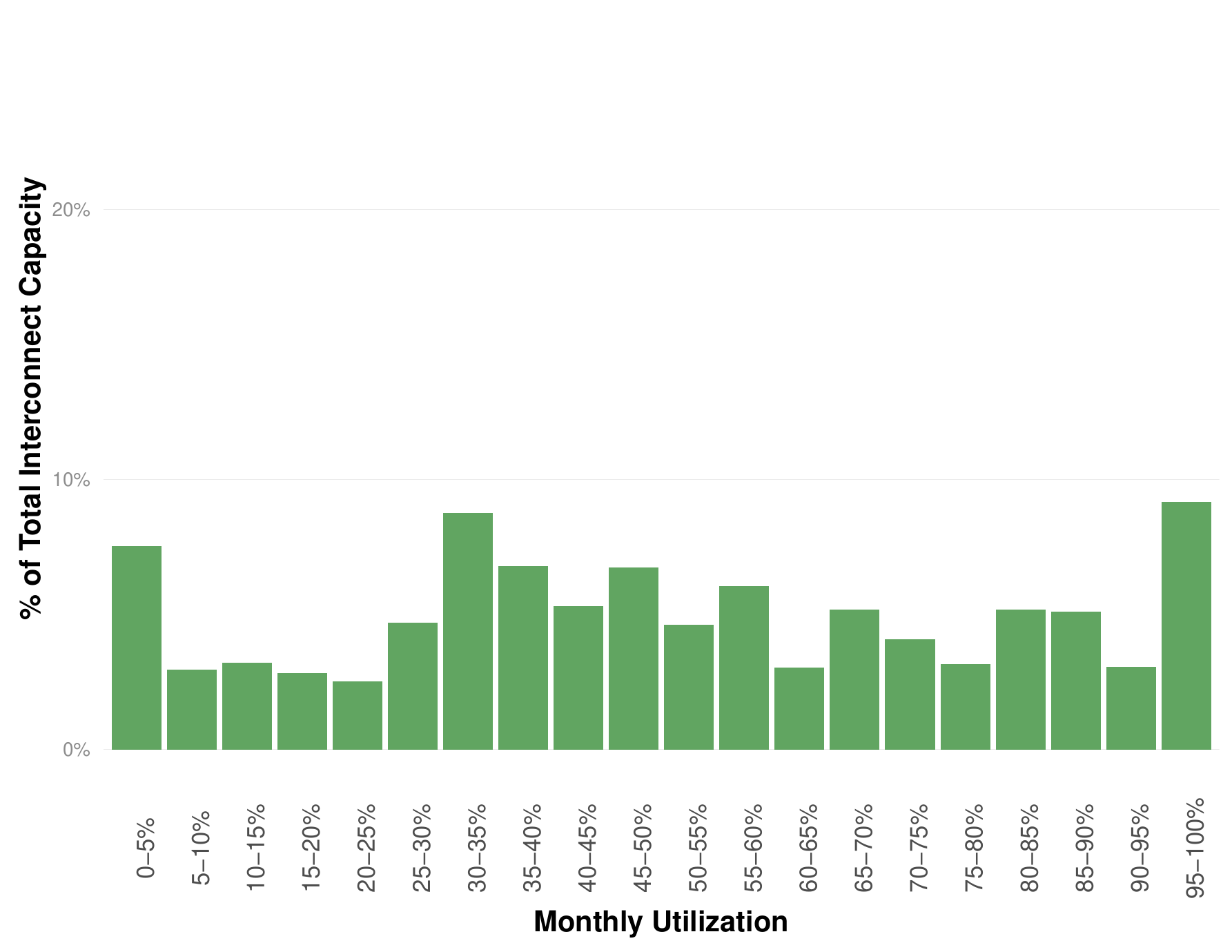}
\caption{Weighted by capacity.\label{fig:dist-cap-peak}}
\end{subfigure}
\end{minipage}
\caption{The fraction of interconnect capacity, weighted by the number
  of links and the amount of total capacity, respectively, whose 95th
  percentile utilization in a month experienced a particular utilization
  level. The figure shows statistics for February 2016.}
\label{fig:dist-peak}
\end{figure}

\begin{figure}[t]
\includegraphics[width=\linewidth]{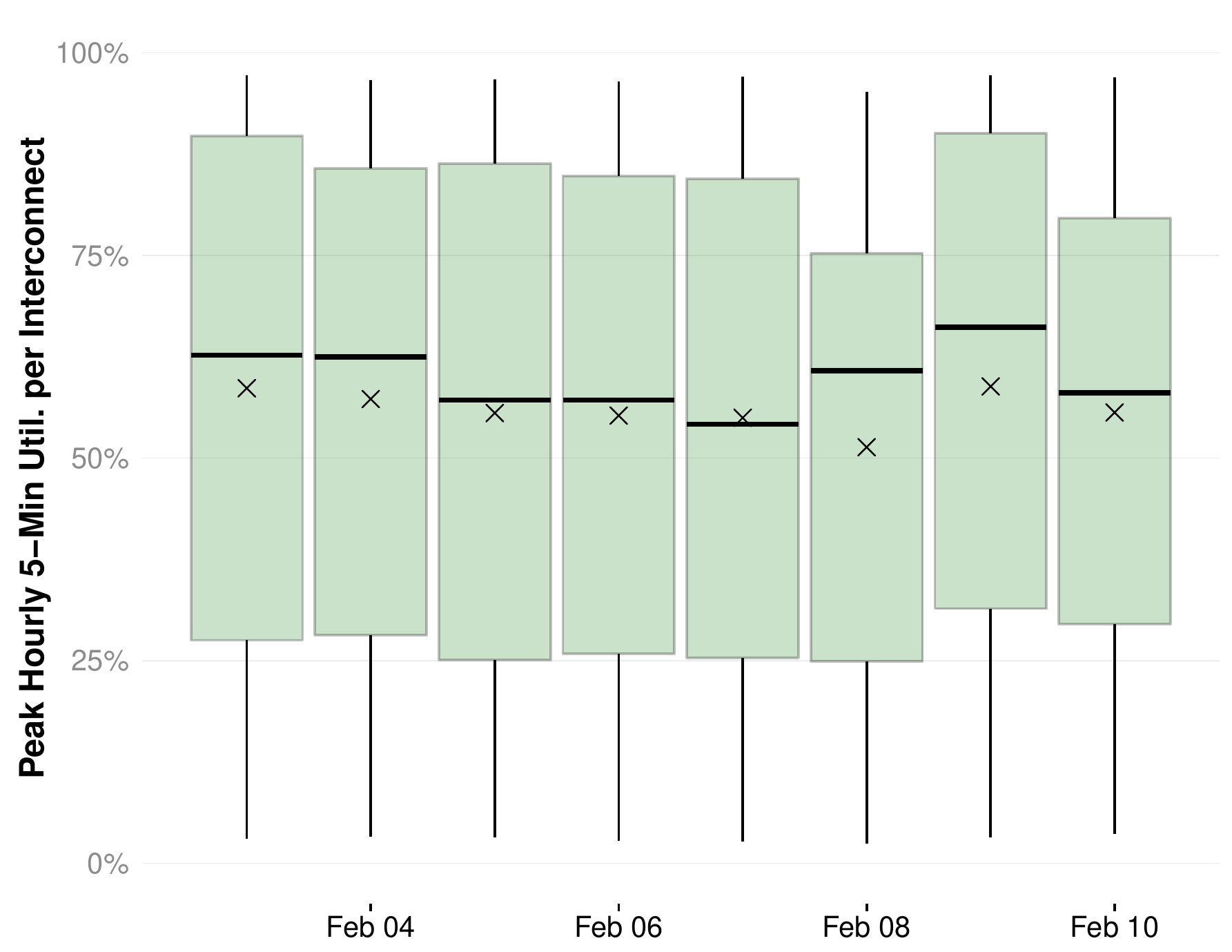}
\caption{Utilization of each interconnect group over one week in
  February 2016 across
  interconnects in Chicago, IL, normalized by
  capacity of the interconnects.} 
\label{fig:utilization-timeseries-chicago}
\end{figure}

\begin{figure}[t!]
\begin{minipage}{1\linewidth}
\begin{subfigure}[b]{\linewidth}
\includegraphics[width=\linewidth]{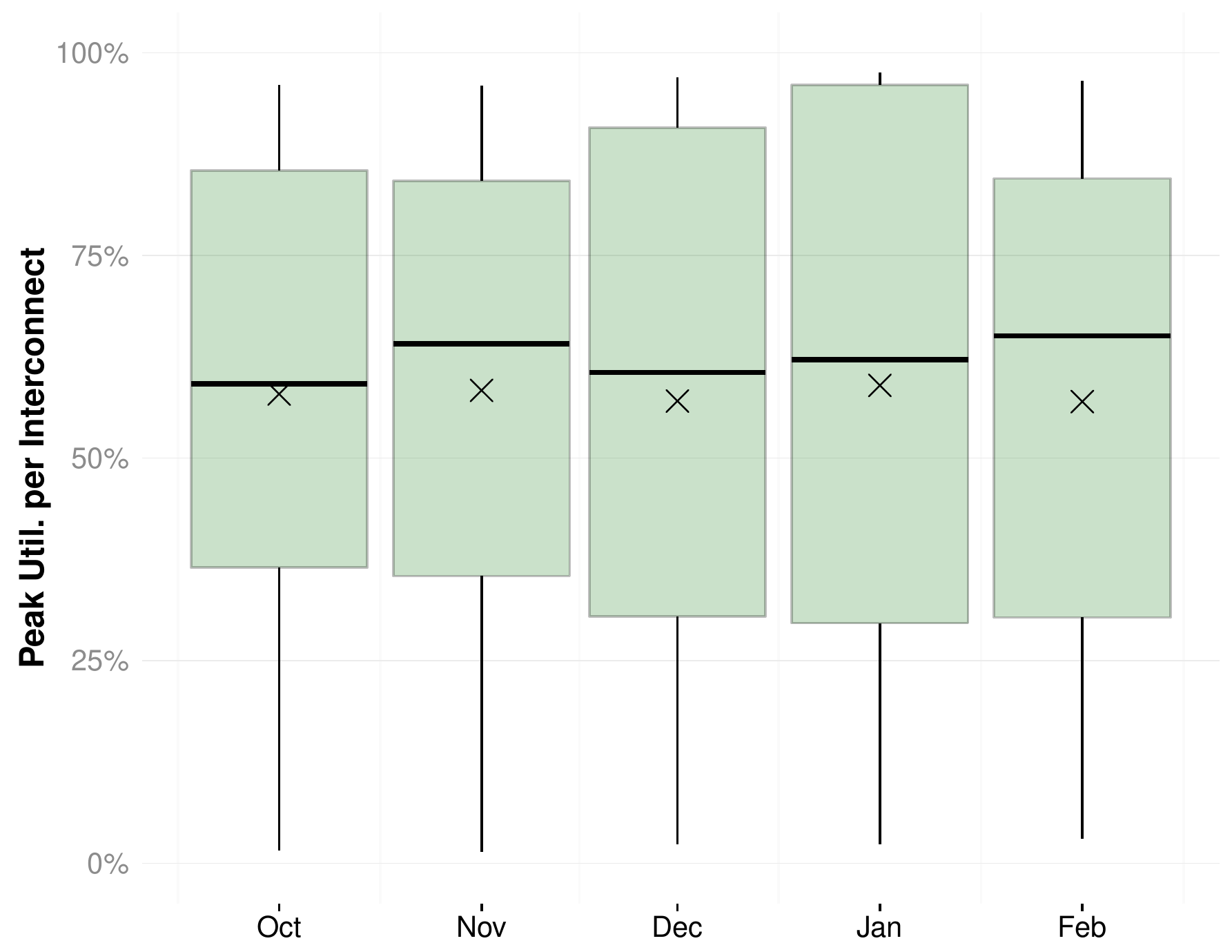}
\caption{Chicago.\label{fig:utilization-per-month-chicago}}
\end{subfigure} \hfill
\begin{subfigure}[b]{\linewidth}
\includegraphics[width=\linewidth]{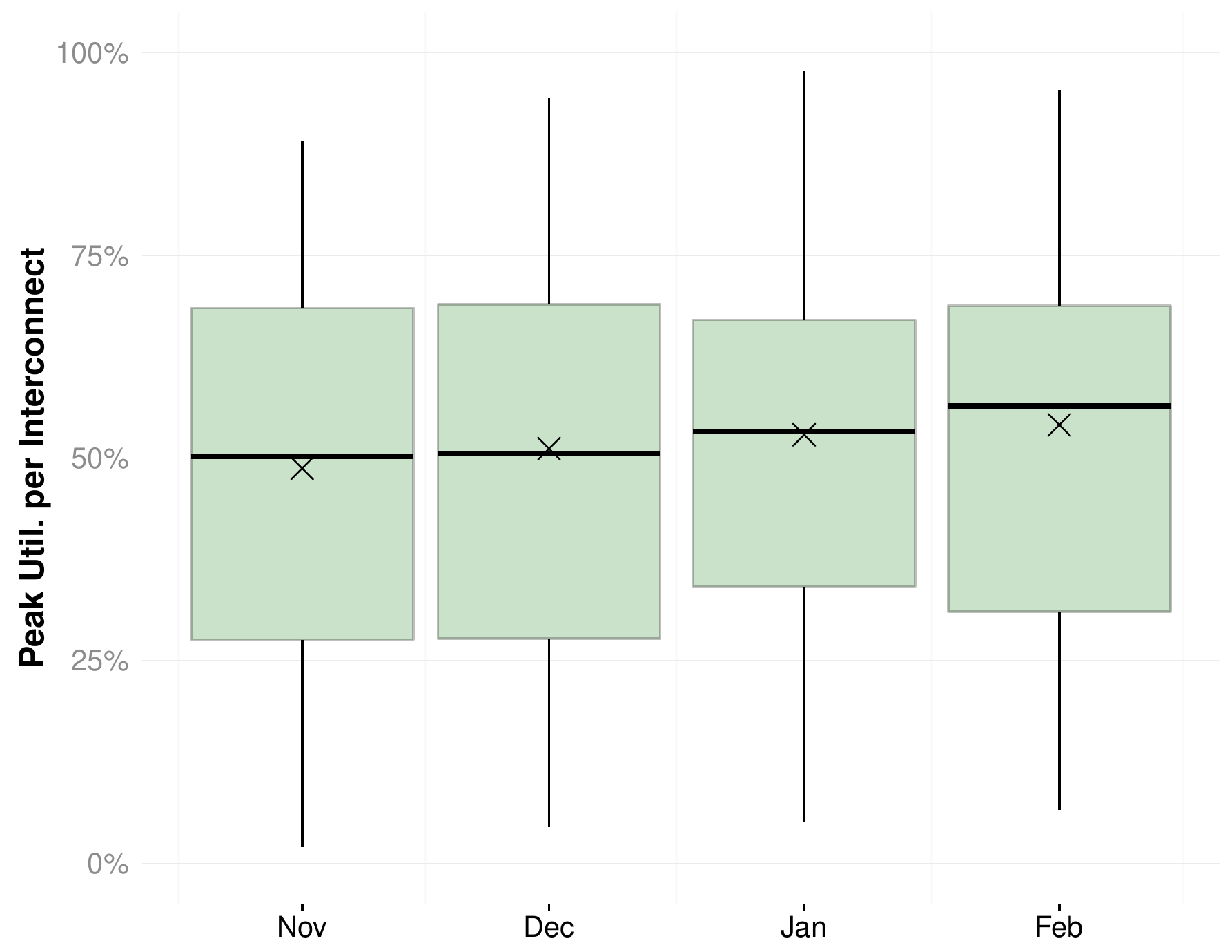}
\caption{San Jose.\label{fig:utilization-per-month-sanjose}}
\end{subfigure}
\end{minipage}
\caption{Per-month utilization of participating interconnects in two
  example regions.}
\label{fig:utilization-per-month-compare}
\end{figure}

\subsection{Aggregate Utilization}

Figure~\ref{fig:utilization-timeseries} shows the
interconnect utilization over time, for a one-week period in
February 2016 across all regions. Each data point in the timeseries
shows a box plot illustrating the distribution of utilization
across interconnect points. The median utilization across interconnects
is consistently below 50\%, even at peak times, and many of the links
have significantly less utilization.  Less than 4\% of the link aggregation groups exceed 95\%
utilization in any five-minute interval, and the vast majority of the
link aggregation groups see much less utilization, even at peak
times. In the next section, we explore these trends for individual
regions. 

Recall that, due to aggregation, we cannot determine whether a
utilization of, say, 75\% indicates that there are no links in the
aggregation group running at full utilization. What we {\em can}
conclude, however, is that there {\em exist} links in the aggregation
group with sufficient spare capacity, and thus that most senders of
traffic have the ability to send traffic flows over links at the
interconnect that have spare capacity, even as other links may have high
utilization. 


Figure~\ref{fig:utilization-per-month-all} shows the distribution of
interconnect capacity by peak utilization over all five-minute intervals across link
aggregation groups for each month, for all aggregation groups. The box
plot shows the inter-quartile ranges, the horizontal line shows the
median utilization, and the whiskers show the 5th and 95th
percentiles. 

\subsection{Utilization by ISPs and Links}

Figure~\ref{fig:isp-dist} shows the distribution of 95th percentile peak ingress
utilization across all ISPs, normalized by capacity. The median ISP in
the group of seven ISPs experienced a 95th percentile peak ingress
utilization that was less typically around 50\% of the available
capacity.  This plot shows that each ISP has significant spare capacity
across its set of links and regions.  This figure does {\em not} indicate
whether a particular ISP is experiencing congestion in a particular
region, to a particular partner network, or across a set of links.

Unfortunately, we cannot show utilization for specific links or neighbor
networks, because the existence of a particular business relationship or
even the existence of a specific link in a region may reveal proprietary
information. We can, however, explore the utilization across the
aggregate of all links, which also shows the existence spare capacity.
Specifically, we can show how the characterization of peak utilization
across {\em all} links, weighted both by links and by overall capacity,
as shown in Figure~\ref{fig:dist-peak}.  Figure~\ref{fig:dist-link-peak}
shows the distribution of 95th percentile peak monthly utilization
across all links, for all participating ISPs.  This figure shows that
more than 25\% of all links are significantly underutilized, and that
less than 10\% of all links experience a 95th percentile peak
utilization that exceeds 90\%.  

In Figure~\ref{fig:dist-link-peak} all links are weighted equally, which
does not reveal whether there exists significant excess {\em capacity},
only whether there exist links that have spare capacity. Exploring
utilization where the set of links is weighted by their capacity reveals
more information. Figure~\ref{fig:dist-cap-peak} shows the same
distribution, where links are weighted by overall capacity. The figure
shows that links that account for about 10\% of overall interconnect
capacity experienced a 95th percentile peak utilization that exceeded
95\%.  Most of the capacity experienced significantly less utilization.

Together, these plots present a picture of the existence of spare
utilization across many of the interconnects that also account for much
of the capacity at interconnects. Certain answers remain obscured, such
as whether a particular partner network is experiencing persistent
congestion, or whether particular types of connections (e.g., paid
peering) are experiencing more or less congestion. Yet, the figures
above do reveal a general picture of (1)~all ISPs having spare capacity
in aggregate across interconnects; (2)~most interconnect capacity in
aggregate showing spare capacity at peak. Both of these conclusions
reveal significantly more than we have known to date; as this project
matures and we receive further feedback, we hope to make additional
views of the data available that also respect the private and
proprietary information of each ISP.

\subsection{Utilization by Region}

We also explored how utilization evolves over time in individual
regions, to determine whether utilization patterns at interconnects in
specific regions agreed with the overall general trends that we observed
in Figure~\ref{fig:utilization-timeseries}.
Figure~\ref{fig:utilization-timeseries-chicago} shows how utilization
evolves over time across interconnects in Chicago; the trends in this
specific region are similar to the overall trends.  The trends are
similar in other cities with busy interconnects; interconnects in
Atlanta show similar distributions.

Washington, New York, Dallas, and Los Angeles exhibit similar
utilization trends, although utilization exceeded 90\% less frequently
than it did in Chicago and Atlanta, the two busiest regions.
Figure~\ref{fig:utilization-per-month-compare} shows the distribution of
interconnect capacity across link aggregation groups over all
five-minute intervals. Figure~\ref{fig:utilization-per-month-chicago}
shows this distribution for a busier Interconnect (Chicago); 
Figure~\ref{fig:utilization-per-month-sanjose} shows the same
distribution for San Jose.
Interconnections in San Jose tend to have
lower median utilizations across link groups, although the highest loaded
link groups at peak time also follow similar trends as those that we
observed in Chicago.

\balance\section{Conclusion and Next Steps}\label{sec:conclusion}

Public discourse surrounding interconnection and congestion begs the need
for better visibility into congestion at interconnection points between
ISPs and content providers.  Unfortunately, the methods that exist for inferring
these statistics from the edge using active probes are
inconclusive---cannot accurately pinpoint congestion at interconnection,
and in many cases they cannot disambiguate congestion that occurs on a
forward path from congestion that occurs on a reverse path.

Ultimately, stronger conclusions require more direct measurements of
utilization at the interconnection points themselves. The public data
collected from ISP interconnection points makes it possible to establish
that spare capacity exists at interconnection points in the aggregate,
and that congestion that is observable at the edge may ultimately
reflect the inefficient use of existing capacity.  Until now, all of
this information has been protected by non-disclosure agreements
between ISPs and neighboring networks. Yet, more informed debate requires better data.  This
paper presents a next step in that direction, based on data from
interconnection points from seven major Internet service providers. 

Our preliminary analysis tells a different story than previous direct
measurement approaches have suggested. Specifically, evidence suggests
that, capacity continues to be provisioned to meet growing
demand and that spare capacity does exist at interconnection points,
even though specific links may be experiencing high utilization. We do not speculate on the reasons behind these usage
patterns, which ultimately derive from content (``edge'') providers'
decisions about where to direct traffic, but the patterns appear to show
clear trends: there exists spare capacity at the interconnection points.

The need to assess metrics that directly affect user experience, such as
application quality or the quality of user experience, 
will ultimately require a much richer dataset than that which is
currently available. For example, more work is needed to understand how
the utilization of a link ultimately affects a customer's quality of
experience for a given application. It may be possible, for example,
that high utilization does not adversely affect customer quality of
experience. Future work may include assessing the correlation between
these network-level traffic statistics and the corresponding quality of
experience for different types of applications. 
\label{lastpage}

\end{sloppypar}

\section*{Acknowledgments}
We thank Steve Bauer, Jennifer Rexford, and Walter
Willinger for their feedback on this paper.  This work was made possible
with support from CableLabs.

\footnotesize
\setlength{\parskip}{-1pt}
\setlength{\itemsep}{-1pt}
\balance\bibliography{paper}
\bibliographystyle{abbrv}
}{
}

\end{document}